\documentclass[twocolumn,nofootinbib,floats,amsmath,amssymb]{revtex4}
\usepackage{graphicx}
\usepackage{amsmath}
\usepackage{float}
\usepackage{amssymb}
\usepackage[all]{xy}
\usepackage{amsfonts}
\usepackage{dcolumn}
\usepackage{bm}
\usepackage{xcolor}

\begin{document}

\title{Cosmic expansion with matter creation and bulk viscosity}
 
\author{V\'ictor H. C\'ardenas$^1$}
\email{victor.cardenas@uv.cl}

\author{Miguel Cruz$^2$}
\email{miguelcruz02@uv.mx}

\author{Samuel Lepe$^3$}
\email{samuel.lepe@pucv.cl}

\affiliation{$^1$Instituto de F\'{\i}sica y Astronom\'ia, Universidad de Valpara\'iso, Gran Breta\~na 1111, Valpara\'iso, Chile \\
$^2$Facultad de F\'{\i}sica, Universidad Veracruzana 91000, Xalapa, Veracruz, M\'exico\\
$^3$Instituto de F\'{\i}sica, Facultad de Ciencias, Pontificia Universidad Cat\'olica de Valpara\'\i so, Avenida Brasil 2950, Valpara\'iso, Chile}

\date{\today}

\begin{abstract}
We explore the cosmological implications at effective level of matter creation effects in a dissipative fluid for a FLRW geometry; we also perform a statistical analysis for this kind of model. By considering an inhomogeneous Ansatz for the particle production rate we obtain that for created matter of dark matter type we can have a quintessence scenario or a future singularity known as little rip; in dependence of the value of a constant parameter, $\eta$, which characterizes the matter production effects. The dimensionless age of this kind of Universe is computed, showing that this number is greater than the standard cosmology value, this is typical of universes with presence of dark energy. The inclusion of baryonic matter is studied. By implementing the construction of the particle production rate for a dissipative fluid by considering two approaches for the expression of the bulk viscous pressure; we find that in Eckart model we have a big rip singularity leading to a catastrophic matter production and in the truncated version of the Israel-Stewart model such rate remains bounded leading to a quintessence scenario. For a non adiabatic dissipative fluid, we obtain a positive temperature and the cosmic expansion obeys the second law of thermodynamics.      
\end{abstract}


\maketitle

\section{Introduction} 

The nature of the late times acceleration of the observable Universe is still far from our actual understanding \cite{accel}. However, this challenge has not been a reason for the community to remain staidly; on the contrary, this has motivated an exhaustive search for models or scenarios beyond General Relativity that attempt to roughly describe the current stage of the Universe in order to elucidate the nature of the catalyst of such accelerated expansion, usually termed as dark energy. As is well known, the cosmological constant approach is a promising scenario, but it has yet to face its own battles. For instance, the origin of the cosmological constant must be at Planck scales but its effects are observed only at cosmological scales (the current accelerated expansion). This difference between scales where the cosmological constant becomes relevant has turned the problem unmanageable, at Planck scales the value of the cosmological constant is greater by approximately 120 orders of magnitude than expected. A reconciliation for the capricious behavior of the cosmological constant at different scales was proposed in Ref. \cite{carlip}, but this description depends on a quantum formulation for the fluctuations of spacetime; nowadays there is no quantum theory of gravity, therefore the description made was at a semiclassical level, this scheme it lacks of a full characterization for these quantum fluctuations and consequently of their cosmic evolution.\\

Another more recent but no less controversial problem of our current Universe is the so-called $H_{0}$ tension, being $H_{0}$ the Hubble constant, this value represents the rate of expansion of the Universe at present time. The problem lies in the discrepancy between the value reported for $H_{0}$ by the Planck collaboration \cite{planck} and the one reported in \cite{tension}. Such difference between both values is not attributable to systematic errors. This has originated for example that schemes where dark energy has the peculiarity of interact with dark matter may be a response to solve this tension among other possible scenarios, that is, we keep looking models beyond standard cosmology, see for instance \cite{interact}. An interesting review on the interacting scheme for the dark sector can be found in \cite{wang}. Other scenarios promote that the $H_{0}$ tension is the result of a tension in the value $T_{0}$ of the cosmic microwave background (CMB) temperature. Despite in some schemes the $H_{0}$ tension can be alleviated, this implies a new paradigm regarding the well established value for the temperature of the CMB \cite{temp}. Other proposals suggest that an enhancement in the geometric description of the Universe is a viable alternative to resolve the aforementioned tension with no need of make use of exotic components or a reformulation of our conception of the gravitational interaction. In Ref. \cite{torsion} was found that the $H_{0}$ tension can be mitigated by considering only a non-vanishing torsional tensor to describe the dark matter sector.\\

In this work we consider the bulk viscous effects plus matter production in the cosmological fluid as an alternative to describe the late times behavior of the observable Universe; it is well known that viscosity effects have a negative contribution in the pressure of the cosmological fluid, therefore such effects lead to accelerated cosmic expansion. On the other hand, in some works it is discussed that matter production is due to the expansion of the Universe \cite{prod1}. For this latter scenario the production of particles is consequence only of the dynamical gravitational background, i.e., the created particles interact only with gravity, therefore their abundance is determined by their masses uniquely. Given the features of the component known as dark matter, this theoretical scenario has been the subject of various tests to explain its nature, despite the fact that dark matter only interacts with gravity, there could be various mechanisms for its production \cite{prod2}. As we will see below, the bulk viscous pressure can be written in terms of the matter production rate if we take into account the adiabatic condition for the entropy, i.e., $S = \mbox{constant}$. We will focus on the cosmological implications of the model at effective level. In the first part of the work we assume an inhomogeneous matter production rate, in general grounds this term is assumed to be constant or can be given as a function of the Hubble parameter $\Gamma \propto H$, but in this work we consider a generalization for this production rate given as $\Gamma = \Gamma(H,\dot{H})$. A similar form for this term was considered in Ref. \cite{us1} by the authors. We obtain that for this kind of model in order to have a consistent description we must include baryonic matter, once it is included, we obtain that a quintessence dark energy behavior is allowed by the model at effective level if we restrict the parameters of the resulting model with the use of recent observational data. In the second part of the work we drop the Ansatz philosophy for the matter production term, $\Gamma$, and we consider two approaches for the dissipative effects in the cosmic fluid, in this second part such approaches lead us to a specific form for the matter production term. As we will see later, in one case the description allows a phantom cosmology leading to a catastrophic production of matter. However, for the second approach the model only allows a quintessence dark energy behavior again. An interesting characteristic of a dissipative fluid is the generation of entropy, in the perfect fluid description the entropy production and heat dissipation are lost \cite{maartens}. However, a recent work showed that at cosmological scales the second law of thermodynamics is still valid \cite{manuel}. On the other hand, in the literature can be found that bulk viscous effects are not ruled out at all by the observational data a provide a framework in which the $H_{0}$ tension can be weakened \cite{andronikos}. As additional examples, see for instance the Ref. \cite{spin}, where was stated that the presence of dissipative effects in the fluid contribute significantly to reproduce the experimental measurements of the longitudinal polarization of hyperons produced in relativistic heavy-ion collisions. In Ref. \cite{cmb} was found that when the inflationary process is attenuated by dissipative causes, the inflaton interchanges its energy density with a emerging radiation component, which can be associated to the CMB and at late times such model leads to a cosmological constant type evolution.\\

Over the years the approach in which matter production effects are employed has changed slightly. At first it was thought that these effects were important only in the early Universe providing a natural explanation for the reheating phase of the inflationary process \cite{reheat} or depending on the rate of matter production, the origin of the Universe could be free of an initial singularity \cite{lima}. However, in the meantime it has also been shown that the matter production effects can play an important role in cosmic evolution, in Ref. \cite{us1}, supported by cosmological observations the authors showed that this scheme leads to an Universe in which the dark energy sector can be emulated by the particle production. Other relevant scenarios in which the production of matter has an important character can be found in \cite{cosmography}, this work explores the possibility that the current state of the Universe is transitory and will eventually present a decelerated phase, this transition is possible by considering matter creation in their model. Another example is given in \cite{entropic} where the cosmological model encloses entropic forces with matter creation. Other interesting works can be found in \cite{lima1, lima2, lima3, lima4}.\\

The outline of this work is as follows: in Section \ref{sec:viscosity} we provide the cosmological equations for a model with matter creation effects and viscosity. Under the adiabatic condition for the entropy we construct the effective parameter state and we explore an inhomogeneous Ansatz for the particle production rate. We obtain the corresponding Hubble parameter and we calculate the age of this kind of Universe and we consider the inclusion of baryons. In Section \ref{sec:observations} we perform the statistical analysis of the model with baryonic matter and the use of current observational data. In Section \ref{sec:constructing} we proceed in the inverse order, with the consideration of the dissipative and matter creation effects we construct the form of the particle production rate, $\Gamma$, we do not use the Ansatz philosophy. We explore the well known Eckart model and the truncated version of the causal Israel-Stewart description, where a cosmological constant evolution can be emulated. In Section \ref{sec:nonadia} we provide a description of the cosmological model when the entropy production is not adiabatic, we find that the temperature of the fluid is positive and obeys the second law of thermodynamics. Under this approach the cosmic expansion does not have the problem of negative entropy or temperature. Finally, in Section \ref{sec:final} we give the final comments of our work. We will consider $8\pi G=c=k_{B}=1$ units throughout this work.  

\section{Matter creation and bulk viscosity}
\label{sec:viscosity}

For a dissipative fluid the local equilibrium scalars such as the particle number density and its energy density are not altered by the dissipative effects. However, the pressure deviates from the local equilibrium pressure
\begin{equation}
    p_{\mathrm{eff}} = p + \Pi,
\end{equation}
where $\Pi$ is the bulk viscous pressure. From now on the quantity, $p+\Pi$, will denote effective pressure for the dissipative fluid. In this description the Friedmann equations for a flat FLRW spacetime can be written as follows
\begin{equation}
    3H^{2} = \rho, \ \ \ \dot{H}+H^{2} = -\frac{1}{6}[(1+3\omega)\rho + 3\Pi],
    \label{eq:fried}
\end{equation}
being $\rho$ the energy density of the dissipative fluid, $H$ denotes the Hubble parameter and the dot stands for derivatives with respect to time. In the previous equations we have considered a barotropic equation of state between the density and the pressure given as, $p = \omega \rho$, where $\omega$ is commonly known as parameter state and it is constrained to the interval $[0,1)$. Note that the consideration of the Friedmann equations given in (\ref{eq:fried}) leads to a non conservation equation for the energy density
\begin{equation}
    \dot{\rho}+3H(1+\omega)\rho = -3H\Pi.
    \label{eq:continuity}
\end{equation}
On the other hand, if matter creation exists, i.e., gravitational particle production, then the continuity equation for the particle number density takes the form 
\begin{equation}
\dot{n} + 3H n = n \Gamma,
\label{eq:number}
\end{equation}
where the possibilities $\Gamma >0$, $\Gamma < 0$ denote source or sink of particles, respectively. $\Gamma$ it is also known as the particle production rate. From the Gibbs equation \cite{maartens}
\begin{equation}
T dS = d \left(\frac{\rho}{n}\right) + p d \left(\frac{1}{n}\right),
\label{eq:gibbseq}
\end{equation}
we can write
\begin{eqnarray}
    nT\dot{S} &=& \dot{\rho} - \rho(1+\omega)\frac{\dot{n}}{n}, \nonumber \\
    &=& - 3H\Pi - \rho \Gamma (1+\omega),
    \label{eq:entropy}
\end{eqnarray}
where the Eqs. (\ref{eq:continuity}) and (\ref{eq:number}) were considered. Note that for this approach the entropy is not longer a constant. However, if we assume $\dot{S}=0$, i.e, an adiabatic dissipative fluid with particle creation in order to be in agreement with the standard cosmological model, we obtain the following condition
\begin{equation}
    \Pi = - \rho(1+\omega)\frac{\Gamma}{3H},
    \label{eq:pi}
\end{equation}
we will have a negative contribution from the viscous pressure to the non equilibrium pressure if the particle production rate is positive, i.e., no annihilation. If we insert the previous equation in (\ref{eq:continuity}) we can write the continuity equation for the density in its standard form, $\dot{\rho} + 3H(1+\omega_{\mathrm{eff}})\rho = 0$, where the effective parameter state has the form
\begin{equation}
    \omega_{\mathrm{eff}} = \omega - (1+\omega)\frac{\Gamma}{3H}.
    \label{eq:effective}
\end{equation}
It is worthy to mention that if the dissipative fluid behaves as standard dark matter we have, $\omega = 0$ ($p=0$), then the effective parameter only depends of the particle production effects, this case was considered in Ref. \cite{us1}, in such case the condition $\Gamma > 3H$ must be fulfilled in order to have an effective phantom behavior.  

\subsection{Ansatz for the particle production rate}

\begin{figure}[htbp!]
\centering
\includegraphics[width=7.8cm,height=6cm]{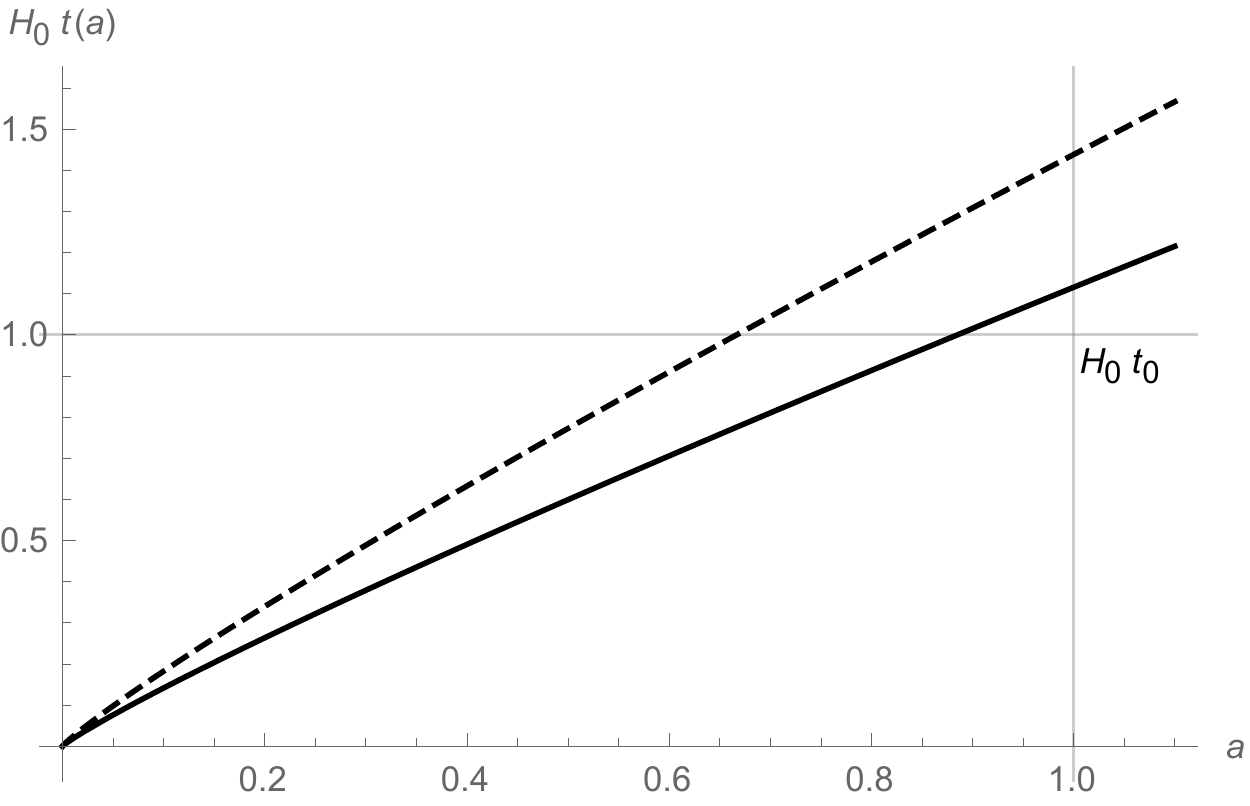}   
\caption{Age of the Universe with quintessence behavior. The dashed line corresponds to $\Omega_{\rho} \approx 0.3$ and the solid line corresponds to $\Omega_{\rho} = 1$. The value $\omega = 1/3$ was considered for both plots.}
\label{fig:age}
\end{figure}

The most simple assumption for $\Gamma$ is given by a constant production rate. However, in order to study the implications of matter production on the cosmic expansion, some specific functions of the Hubble parameter for the particle production rate have been studied, for instance $\Gamma \propto H^{\alpha}$, being $\alpha$ an appropriate constant \cite{us1, generalrate, generalrate1, Nunes:2015rea, victor, victor1}. A more general form for $\Gamma$ can be found in Ref. \cite{Nojiri:2005sr}, where
\begin{equation}
\Gamma = \Gamma(\rho, p(\rho), H, \dot{H}, \ddot{H},...),
\label{eq:inhomo}
\end{equation}
and the function $p(\rho)$ is viable by means of the equation state. The form given in Eq. (\ref{eq:inhomo}) for $\Gamma$ it is known as inhomogeneous particle production rate. In our case we will consider $\Gamma$ given as
\begin{equation}
    \Gamma = \Gamma(H,\dot{H}) = \gamma'(H)\dot{H} = \frac{d\gamma(H)}{dt},
    \label{eq:inhomo2}
\end{equation}
where the prime denotes derivatives with respect to the Hubble parameter by means of the chain rule. Note that the form of $\Gamma$ given above together with Eq. (\ref{eq:pi}) lead us to a direct integration integration of Eq. (\ref{eq:continuity}), yielding
\begin{equation}
    \rho = \rho_0 a^{-3(1+\omega)} \exp{[(1+\omega)\gamma(H)]},
\end{equation}
then if $\gamma(H) \propto \ln (H^{\delta})$ and $\delta$ being a constant, we could write consistently an expression for the Hubble parameter by means of the first Friedmann equation. We focus on the following expression
\begin{equation}
    \gamma(H) = 2\left(1-\frac{1}{\eta}\right)\ln \left(\frac{H}{H_{0}}\right),
    \label{eq:gamma}
\end{equation}
being $\eta$ a constant value and $H_{0}$ the value of the Hubble parameter at present time. Using the above $\gamma$ function and (\ref{eq:inhomo2}), we can write
\begin{equation}
    \frac{\Gamma}{3H} = -\frac{2}{3}\left(1-\frac{1}{\eta} \right)(1+q),
    \label{eq:quotientnonadia}
\end{equation}
where $q$ is the deceleration parameter defined as $1+q := -\dot{H}/H^{2}$. As can be seen, the particle production rate can be written as a function of the deceleration parameter. On the other hand, using the acceleration equation (\ref{eq:fried}) together with the expression for the viscous pressure obtained from the adiabatic condition (\ref{eq:pi}), one gets
\begin{equation}
    \frac{\Gamma}{3H} = 1 -\frac{2}{3}\frac{(1+q)}{(1+\omega)},
    \label{eq:quotient}
\end{equation}
by equating both expressions for the quotient $\Gamma/3H$, we arrive to the following result
\begin{equation}
    1+q = \frac{3}{2}\left\lbrace \frac{1}{1+\omega} - \left(1-\frac{1}{\eta}\right)\right\rbrace^{-1},
    \label{eq:decelconst}
\end{equation}
then, the deceleration parameter takes a constant value if matter creation effects are introduced in a dissipative fluid. Note that if in the previous expression we consider the value, $\eta = 2(1+\omega)/(3+5\omega)$, we have $q=0$, which represents a Dirac-Milne universe. This kind of Universe expands at constant rate since $\ddot{a} = 0$. For standard dark matter we have, $q = - (1-3\eta/2)$, in this case the accelerated cosmic expansion will take place only if the condition $\eta < 2/3$ is satisfied and no accelerated expansion for $\eta = 2/3$. Using the Eq. (\ref{eq:quotient}) we can write for the effective parameter (\ref{eq:effective})
\begin{equation}
    \omega_{\mathrm{eff}} = -1 + \frac{2}{3}(1+q) = -1 + \left\lbrace \frac{1}{1+\omega} - \left(1-\frac{1}{\eta}\right)\right\rbrace^{-1}.
    \label{eq:omegaeff}
\end{equation}
As expected, the effective parameter state has a contribution from the matter creation effects. If the following conditions are satisfied $\eta > (1+\omega)/\omega$ or $0 < \eta < 2(1+\omega)/(3+5\omega)$, the effective parameter (\ref{eq:omegaeff}) will behave as a phantom or quintessence fluid, therefore accelerated cosmic expansion can be obtained for a dissipative fluid with matter creation effects in presence of ordinary matter. On the other hand, for $\omega = 0$ we have $\omega_{\mathrm{eff}} = -1 + \eta$, thus from the condition $0 < \eta < 2/3$, the dissipative fluid will behave as quintessence and will exhibit phantom behavior with $\eta < 0$. For a Dirac-Milne Universe we will have $\omega_{\mathrm{eff}} = -1/3$.\\ 
Using the Friedmann constraint (\ref{eq:fried}) we have, $3H^{2} = \rho = \rho_{0}a^{-3(1+\omega)}\exp(\gamma(H))$, rearranging properly the terms appearing in this expression after inserting the Eq. (\ref{eq:gamma}); we obtain for normalized Hubble parameter
\begin{equation}
    E(z) = \Omega_{\rho}^{1/2(1-\Delta)}(1+z)^{3(1+\omega)/2(1-\Delta)},
    \label{eq:hubble}
\end{equation}
where we have defined $\Delta := (1+\omega)(1 - 1/\eta)$, $\Omega_{\rho}$ corresponds to the standard definition of the density parameter, $\Omega_{\rho} := \rho_{0}/3H^{2}_{0}$ and we also used the relation between the scale factor and the redshift, $1+z = a^{-1}$, besides $E(z) := H(z)/H_{0}$. Note that for $\eta = 1$ we have a null contribution from matter creation effects and we recover the standard cosmology. In the standard dark matter case we have
\begin{equation}
    E(z) = \Omega_{\rho}^{1/2(1-\Delta)}(1+z)^{3/2(1-\Delta)},
\end{equation}
for the quintessence scenario we have $0 < \eta < 2/3$, therefore $ - \infty < \Delta < - 1/2$, then as we approach to the far future, we have for the normalized Hubble parameter, $E(z\rightarrow -1) \rightarrow 0$. For the phantom regime we have $\eta < 0$, thus $\Delta > 1$, which leads to a singular scenario as we approach to the far future, $E(z\rightarrow -1) \rightarrow \infty$. A similar behavior is obtained for the energy density as we approach to the far future, for quintessence $\rho \rightarrow 0$ and $\rho \rightarrow \infty$ for phantom. It is worthy to mention that in the phantom scenario the singular nature takes place only at the far future and not for a specific value of the redshift, therefore this kind of singularity corresponds to a little rip. From the current observational data it is not possible to determine if the final fate of the Universe is a future singularity or not but if consistency with the supernova is demanded, then this kind of singular model could represent a viable alternative to the $\Lambda$CDM model \cite{little1, little2, little3}. If we consider the Eq. (\ref{eq:hubble}) and $H(a) = d\ln a/dt$, we can compute the dimensionless age of the Universe
\begin{equation}
    H_{0}t(a) = \int^{a}_{0} \frac{da'}{a'H(a')},
\end{equation}
as stated in Ref. \cite{avelino}, the value for the dimensionless age at present time, $H_{0}t_{0}$, is around 1 for the $\Lambda$CDM model and this is independent of the transition from the decelerated to accelerated stage, the authors call this fact as {\it synchronicity problem} given that it appears we are living in a special time. In Fig. (\ref{fig:age}) we show the behavior of the dimensionless age of the Universe if we consider the Eq. (\ref{eq:hubble}) with a quintessence behavior. We also consider the value $\omega = 1/3$ and we give the value $0.315 \pm 0.007$ \cite{planck} to the energy density parameter $\Omega_{\rho}$, according to the latest Planck collaboration results, this case corresponds to the dashed line in the figure. On the other hand, if we consider that created matter is the only component of the Universe we must have, $\Omega_{\rho} = 1$, this corresponds to the solid line of the plot, this latter case is closer to the value 1 at present time ($a=1$). Note that at present time the value of the dimensionless age of the Universe is increased in both cases, however, the obtained values are close to 1; the augmentation in the age of the Universe at present time by dark energy was discussed in Ref. \cite{caldwell} for phantom dark energy.  

\subsection{Including baryonic matter}

It is worthy to mention that if we consider only the matter creation effects we are left with a constant deceleration parameter in our description, this can be seen in Eq. (\ref{eq:decelconst}). Despite that dark energy behavior can be obtained for certain values of the parameters $\eta$ and $\omega$, the model itself is not consistent with the $\Lambda$CDM model and the type Ia supernovae data; this issue can be alleviated with the inclusion of baryons as done in Ref. \cite{baryons}. In this new scenario the form of the normalized Hubble parameter can be expressed as follows
\begin{equation}
    \left(\frac{H}{H_{0}} \right)^{2} = \Omega_{B}(1+z)^{3} + (1-\Omega_{B})(1+z)^{3(1+\omega)}\left(\frac{H}{H_{0}} \right)^{2\Delta},
    \label{eq:hubble2}
\end{equation}
where we take into account the normalization condition, $\Omega_{B} + \Omega_{CM} = 1$, i.e., for the late times description of the Universe we only consider the contribution from the gravitationally created matter and baryons. From the previous expression it is not possible to find the normalized Hubble parameter analytically. However, if we consider the acceleration equation (\ref{eq:fried}) with the assumption, $p_{B} = 0$, and the equations (\ref{eq:pi}), (\ref{eq:gamma}) together with the standard relationship between the redshift and the scale factor given above; we can write the following differential equation for the normalized Hubble parameter
\begin{widetext}
\begin{equation}\label{eq20}
   (1+z)E\frac{dE}{dz}-E^{2}-\frac{1}{3}(1+\omega)(1-\Omega_{B})(1+z)\left(1-\frac{1}{\eta}\right)\frac{d \ln E}{dz} = \frac{1}{2}\left[(1+3\omega)(1-\Omega_{B})+\Omega_{B}(1+z)^{3}\right],
\end{equation}
\end{widetext}
from which $E(z)$ can be obtained by numerical integration. We will use this equation to constraint the parameters of the model with the use of current cosmological observations later. In this we will have at effective level from the contribution of baryons and matter creation effects
\begin{equation}
    \omega_{\mathrm{eff}} = \frac{\eta - 1}{1+\frac{\Omega_{B}}{3H^{2}_{0}(1-\Omega_{B})^{\eta}}(1+z)^{3(1-\eta)}},
    \label{eq:omegabar}
\end{equation}
where we have considered the equation (\ref{eq:pi}) for the pressure together with the assumption that created matter behaves as standard dark matter, $\omega=0$, and $p_{B}=0$.

\section{Observational constraints}
\label{sec:observations}

\begin{figure}[htbp!]
\centering
\includegraphics[width=7cm,height=7cm]{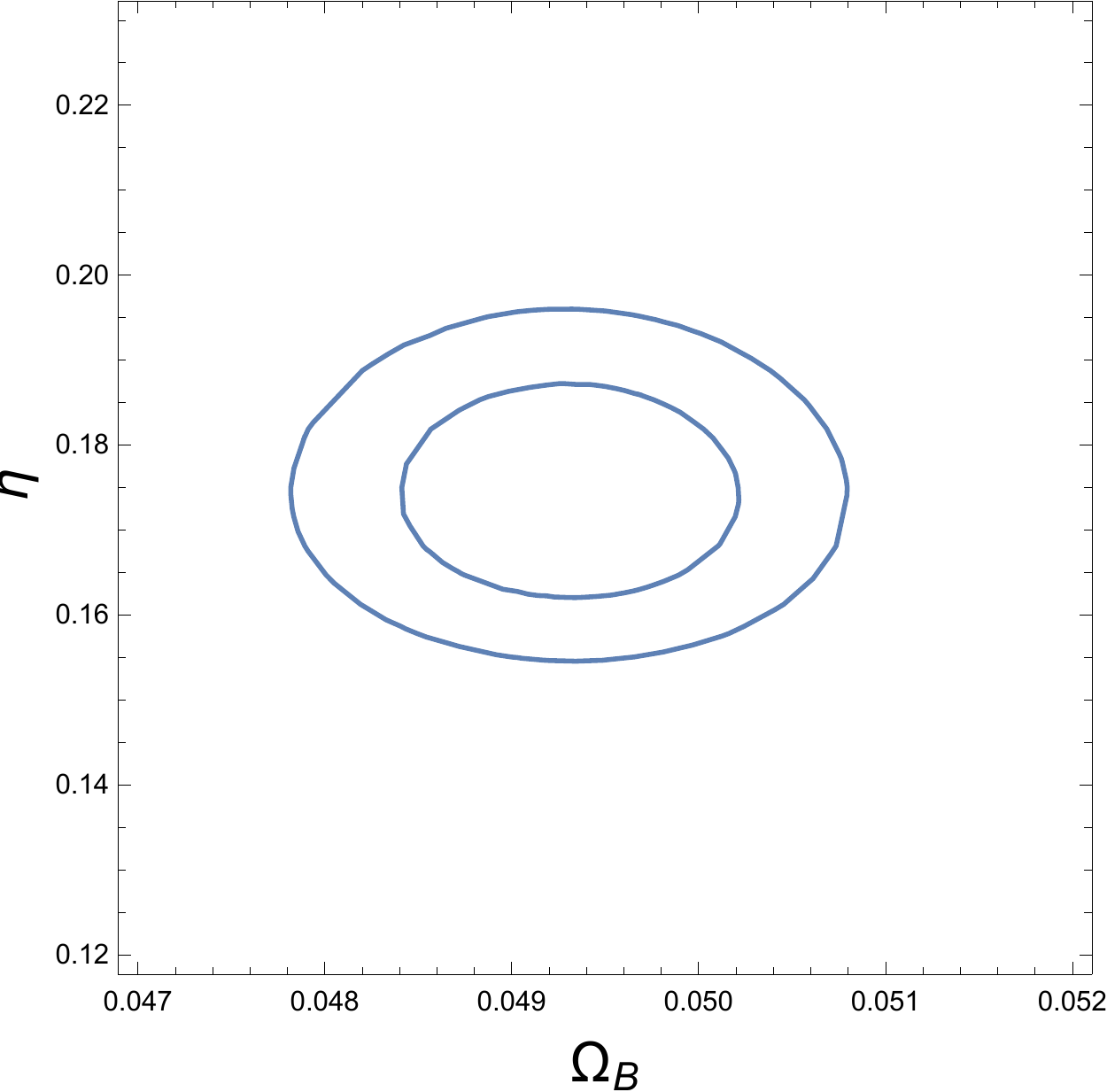}   
\caption{Confidence contours at $1\sigma$ and $2\sigma$ for the $\eta$ and $\Omega_B$ parameter in the case of Eq. (\ref{eq20}) keeping $w=0$.}
\label{fig:etavsomb}
\end{figure}

In this section we study to what extend the evolution implied by Eq. (\ref{eq20}) describes appropriately the observations. In particular, we use the latest type Ia supernovae sample called Pantheon \cite{panteon} consisting in 1048 data points. The data gives us the apparent magnitude $m$ at maximum from which we can compute the distance modulus $\mu = m-M$ with $M$ the absolute magnitude for type Ia supernovae. Here we compute the residuals $\mu - \mu_{th}$ and minimize the quantity
\begin{equation}\label{chi2jla}
\chi^2 = (\mu - \mu_{th})^{T} C^{-1}(\mu - \mu_{th}),
\end{equation}
where $\mu_{th} = 5 \log_{10} \left( d_L(z)/10pc\right) $ gives the theoretical distance modulus, $d_L(z)$ is the luminosity distance given by
\begin{equation}
    d_L(z)= (1+z)\frac{C}{H_0}\int \frac{dz}{E(z)},
\end{equation}
$ C $ is the covariance matrix released in \cite{panteon}, and the observational distance modulus takes the form
\begin{equation}\label{mujla}
\mu = m - M + \alpha_1 X - \alpha_2 Y,
\end{equation}
where $ m $ is the maximum apparent magnitude in band B, $ X $ is related to the widening of the light curves, and $ Y $ corrects the color. usually, the cosmology -- specified here by $\mu_{th}$ -- is constrained along with the parameters $ M $, $\alpha_1 $ and $\alpha_2$. These nuisance parameters are then marginalized to obtain the posterior probabilities for our parameters of interest: $w$, $\Omega_B$ and $\eta$. In the statistical study we use a prior for $\Omega_B$ and we assume $w=0$ assuming the contribution be as dark matter. Then, the free parameters to fit are $\Omega_B$ and $\eta$. Using the value $\Omega_B=0.0493 \pm 0.0006$ based on \cite{PDG}. In this case it is possible to obtain as best fit values $\Omega_B = 0.049 \pm 0.001$, and $\eta=0.174 \pm 0.015$. The confidence contours for $1\sigma$ and $2\sigma$ are shown in Fig. (\ref{fig:etavsomb}). If we insert the constrained values for $\Omega_{B}$ and $\eta$ in Eq. (\ref{eq:omegabar}) together with the Hubble constant reported in Ref. \cite{planck} one gets that at present time the effective parameter lies in the interval $[-0.840997,-0.810997]$; this interval is within the quintessence region determined by the DES collaboration (Dark Energy Survey) for the parameter state at present time, $\omega _{\mathrm{de,0}} = -0.95^{+0.33}_{-0.39}$ \cite{des}. On the other hand, as we approach the far future, from Eq. (\ref{eq:omegabar}) one gets $\omega_{\mathrm{eff}}(z \rightarrow -1) \rightarrow \eta -1 = -0.826 \pm 0.015$, thus the value of the parameter state at the far future is very close to its present time value. Therefore this model represents a quintessence dark energy behavior along the cosmic evolution and is in agreement with some sets of observational data such as the latest Planck results ($\omega_{\mathrm{de,0}}= -1.028 \pm 0.031$) and DES collaboration, where the quintessence behavior for dark energy is allowed.   

\section{Constructing $\Gamma$ from viscous models}
\label{sec:constructing}

In this section we construct the particle production rate from bulk viscous considerations. We will not consider a specific Ansatz for this term. According to the Israel-Stewart formalism, the bulk viscous pressure must obey the following transport equation \cite{is1, hiscock}
\begin{equation}
    \Pi = -3\xi(\rho)H - \tau \dot{\Pi} - \frac{\zeta}{2}\tau \Pi \left[3H + \frac{\dot{\tau}}{\tau} - \frac{\dot{\xi}}{\xi} - \frac{\dot{T}}{T}\right],
    \label{eq:transport}
\end{equation}
where $\xi(\rho)$ is the bulk viscosity coefficient and given that it is a function of the energy density we must have $\xi(\rho) \geq 0$ a typical form for the bulk viscosity coefficient can be found in the literature as $\xi \propto \rho^{s}$ being $s$ a constant; $T$ is the barotropic temperature which is also a function of the energy density by means of the integrability Gibbs condition. In general, an interesting parametrization for the bulk viscosity coefficient is given in terms of the scalar expansion of the fluid, i.e, $\xi \propto \Theta$ where $\Theta = H$, thus $\xi \propto \rho^{1/2}$, i.e., $s=1/2$; physically this represents an increasing viscosity in the case of increasingly movements in the fluid \cite{brevik}\footnote{The introduction of a 4-velocity field $u^{\alpha}$ provides the 1+3 decomposition of spacetime. The metric tensor splits as $g_{\alpha \beta} = h_{\alpha \beta} - u_{\alpha}u_{\beta}$, where $h_{\alpha \beta}$ is the projection tensor and is orthogonal to $u^{\alpha}$. The kinematics of the 4-velocity field takes the form \cite{maartens}
\begin{equation*}
    \nabla_{\alpha}u_{\beta} = \frac{1}{3}\Theta h_{\alpha \beta} + \sigma_{\alpha \beta} + \omega_{\alpha \beta} - \dot{u}_{\alpha}u_{\beta}. 
\end{equation*}
where $\Theta = H$ generalizes the Newtonian expansion and $\sigma_{\alpha \beta}$ and $\omega_{\alpha \beta}$ generalize the Newtonian shear and Newtonian vorticity, respectively. $\dot{u}_{\alpha}$ represents a 4-acceleration (the dot represents a time derivative). Spatial homogeneity and isotropy demand $\dot{u}_{\alpha} = \sigma_{\alpha \beta} =  \omega_{\alpha \beta} = 0$. The scalar expansion determines the volume evolution.}. On the other hand, $\tau$ represents the relaxation time for bulk viscous effects. In Ref. \cite{lindblom} was shown with a perturbative method that the Israel-Stewart theory fulfills the causality and stability conditions. The energy functional can be written as $\sum_{A=1}^{8}\Omega_{A}(\delta Z_{A})^{2}$ where $\delta Z_{A}$ represents certain combination of the perturbation functions, then the positivity of this functional is guaranteed for $\Omega_{A} \geq 0$. A specific term is given as
\begin{equation}
    \Omega_{3}(\lambda) := (\rho + p)\left\lbrace 1- \lambda^{2}\left[\left(\frac{\partial p}{\partial \rho}\right)_{S} + \frac{\xi}{\tau(\rho + p)}\right]\right\rbrace \geq 0, 
\end{equation}
this requirement was shown to hold for all $\lambda$ where, $0 \leq \lambda \leq 1$, by taking the case $\lambda = 1$ one gets
\begin{equation}
    \left(\frac{\partial p}{\partial \rho}\right)_{S} + \frac{\xi}{\tau(\rho + p)} \leq 1, 
    \label{eq:pert}
\end{equation}
note that the first term corresponds to the adiabatic speed of sound, usually denoted as $c^{2}_{s}$, and the second term is identified as the speed of the bulk viscous perturbations, $c^{2}_{b}$, therefore from the above condition we have $v^{2} = c^{2}_{s} + c^{2}_{b} \leq 1$ and this latter expression guarantees the causality of Israel-Stewart model. For a barotropic fluid, $p=\omega \rho$, and using the Eq. (\ref{eq:pert}) we obtain 
\begin{equation}
c^{2}_{b} :=  \frac{\xi}{\tau\rho(1 + \omega)} \leq 1 - \omega,   
\end{equation}
usually the r.h.s. of the previous equation is written as $\epsilon(1-\omega)$, where $\epsilon$ is a constant parameter. Therefore, from the previous equation we write for the relaxation time $\tau = \xi/c^{2}_{b}\rho(1 + \omega) = \xi/\epsilon(1-\omega^{2})$. From the Friedmann equations (\ref{eq:fried}), we can write for the bulk viscous pressure
\begin{equation}
    \Pi = -2\dot{H} - 3(1 + \omega)H^{2}.
    \label{eq:viscpress}
\end{equation}
In order to illustrate some results, in the following sections we will study two cases separately.

\subsection{Eckart model}

If we consider $\tau = 0$ in Eq. (\ref{eq:transport}) we obtain the Eckart model. This approach have been studied exhaustively despite its superluminal propagation problem. However, represents a manageable framework for viscous effects, see for instance the references contained in \cite{eckart}, where the Eckart framework is studied in the late and early Universe and tested with recent observational data. From the Eq. (\ref{eq:transport}) with $\tau = 0$, the Friedmann equation (\ref{eq:fried}) and the standard definition for the bulk viscous coefficient we can write 
\begin{equation}
    \Pi = -3\xi(\rho)H = -3^{s+1}\xi_{0}H^{2s+1},
\end{equation}
if we consider the previous expression in equation (\ref{eq:pi}) we arrive to an explicit form for the particle production rate given as
\begin{equation}
    \frac{\Gamma(H)}{3H} = \frac{3^{s}\xi_{0}}{(1+\omega)}H^{2(s-1/2)},
    \label{eq:rate}
\end{equation}
note that in order to maintain gravitational particle production, $\Gamma > 0$, we must have $\xi_{0} > 0$ for null viscous effects we have not particle production. The case $s=1/2$ simplifies the above equation leading to a constant rate for the production of particles and it is given as $\Gamma = \sqrt{3}\xi_{0}/(1+\omega)$ or $\Gamma = \sqrt{3}\xi_{0}$ for the standard dark matter case. Now, using the Eq. (\ref{eq:pi}) and (\ref{eq:viscpress}) we can write
\begin{equation}
    \frac{\Gamma}{3H} = 1+\frac{2}{3(1+\omega)}\left(\frac{\dot{H}}{H^{2}}\right),
    \label{eq:rate2}
\end{equation}
and by means of (\ref{eq:rate}) the previous expression takes the following form
\begin{equation}
    \frac{\dot{H}}{H^{2}} = -\frac{3}{2} \left\lbrace 1 + \omega - 3^{s}\xi_{0}H^{2(s-1/2)}\right\rbrace.
    \label{eq:diff}
\end{equation}
If we consider the standard dark matter case, $\omega = 0$, together with the case $s=1/2$, then the Eq. (\ref{eq:diff}) can be integrated straightforwardly, yielding
\begin{equation}
H(t) = H_{0}\left[1 + \frac{3\sqrt{3}}{2}\left(\frac{1}{\sqrt{3}} - \xi_{0}\right)H_{0}(t - t_{0}) \right]^{-1},
\end{equation}
where $H_{0}$ denotes evaluation of $H(t)$ at $t = t_{0}$. The expansion of the Universe is guaranteed for $0 < \xi_{0} < 1/\sqrt{3}$. However, for $\xi_{0} > 1/\sqrt{3}$ the Hubble parameter becomes negative, therefore we can write it in the following convenient form
\begin{equation}
    H(t) = \frac{2}{3\sqrt{3}}\left(\frac{1}{\xi_{0}-1/\sqrt{3}} \right)(t_{s}-t)^{-1}, 
    \label{eq:hsing}
\end{equation}
where we have defined
\begin{equation}
    t_{s} = t_{0} + \frac{1}{H_{0}}\left[\frac{2}{3\sqrt{3}(\xi_{0}-1/\sqrt{3})} \right].
\end{equation}
Notice that $t_{s}$ is a constant value and will represent a value for the time at the future at which the Hubble parameter becomes singular. According to the classification for future singularities given in \cite{odintsov1}, we have a big rip singularity at $t = t_{s}$. Thus inserting the Hubble parameter (\ref{eq:hsing}) in Eq. (\ref{eq:rate}) with $s=1/2$ and $\omega = 0$, one gets
\begin{equation}
    \Gamma(t) = 3\sqrt{3}\xi_{0}H = \left(\frac{2\xi_{0}}{\xi_{0}-1/\sqrt{3}} \right)(t_{s}-t)^{-1}.
\end{equation}
As can be seen, the particle production rate becomes singular as we approach to $t_{s}$. This result is contradictory since in the phantom scenario as the Universe reaches the future singularity, matter or any structure must be disaggregated. However, we must keep in mind that have been shown that phantom scenarios are unstable, i.e., at quantum level they have an unbounded negative energy that leads to the absence of a stable vacuum state. On the other hand, in order to avoid this kind of problem in Ref. \cite{phproduction} was proposed the interaction of phantom particles and standard matter, at least gravitationally. In consequence, the gravitational interaction allows processes as the spontaneous production from the vacuum of a pair of phantom particles and a pair of photons, to mention some. In this case the phase space integral is divergent, this indicates a catastrophic instability. This instability can be avoided by imposing a non Lorentz invariant momentum space cutoff, but despite this correction the density number of photons and phantom particles can be written as, $n \propto (\Gamma \times \mbox{age \ of \ the \ Universe})$, and according to the diffuse gamma ray background observations the production rate of photons, $\Gamma$, leads to higher values from the typical ones for the energy of the produced photons. Then, any cosmic ray experiment at earth should be detecting more events than normal and more energetic than those detected until now. These bounds for $\Gamma$ and $n$ together with other considerations of Ref. \cite{phproduction} suggest that the origin of phantom must come from new physics beyond the standard model of particles. See also Refs. \cite{phproduction1, phproduction2} for similar discussions on the topic. 

\subsection{Israel-Stewart approach}

Setting $\zeta = 0$ in the transport equation (\ref{eq:transport}) leads to the truncated version of the Israel - Stewart theory \cite{is1, hiscock}. This effective model has been widely studied at cosmological level in several contexts, see the references in \cite{truncated}. Therefore, if we consider the truncated transport equation (\ref{eq:transport}) together with the Friedmann equations (\ref{eq:fried}) and the continuity equation for the energy density (\ref{eq:continuity}), we obtain a second order differential equation for the Hubble parameter
\begin{align}
& \ddot{H}+ 3H\dot{H}(1+\omega)+\frac{9}{2}\epsilon \left(1-\omega^{2} \right)\left\lbrace \frac{(1+\omega)}{3^{s}\xi_{0}}H^{1-2s} - 1 \right\rbrace H^{3}\nonumber \\ 
& + \frac{\epsilon (1-\omega^{2})}{3^{s-1}\xi_{0}}\dot{H}H^{2(1-s)}=0.
\label{eq:is}
\end{align}
Taking the value $s=1/2$, as discussed in Ref. \cite{svalue}, if dissipative effects are included in the $\Lambda$CDM model, the case $s=1/2$ is the only value that allows to write the standard form of the well known de Sitter solutions, $H_{0} = \pm \sqrt{\Lambda/3}$, in the limit $\xi \rightarrow 0$; and if we also consider the change of variable $x = \ln (a/a_{0})$, we can solve for the Hubble parameter in the standard dark matter case
\begin{equation}
    H(z) = H_{0}\sqrt{\frac{\cos \left(\beta \left[c_{2} + \ln \left(1+z\right) \right] \right)}{\cos (\beta c_{2})}}\left(1+z\right)^{\alpha},
    \label{eq:his}
\end{equation}
where $c_{2}$ is an integration constant given as 
\begin{equation}
    c_{2} = \frac{1}{\beta}\arctan\left(\frac{2\left\lbrace \alpha - (1+q_{0})\right\rbrace}{\beta} \right).
\end{equation}
The form written in Eq. (\ref{eq:his}) for the Hubble parameter is obtained from the condition $H(z=0)$ and the first derivative $H'(z=0)$ given that we are solving a second order differential equation, for simplicity in the notation we have defined $\alpha = 3/4 + \sqrt{3}\epsilon/4\xi_{0}$ together with $\beta = 1/2\sqrt{6\epsilon/\xi_{0}(\sqrt{3}-6\xi_{0}-3\xi_{0}/2\epsilon - \epsilon/2\xi_{0})}$, which are also constants. On the other hand, $q_{0}$ is the deceleration parameter evaluated at $z=0$. Note that at present time $z=0$, the Hubble parameter (\ref{eq:his}) takes the value $H_{0}$.\\

\begin{figure}[htbp!]
\centering
\includegraphics[width=7.8cm,height=6cm]{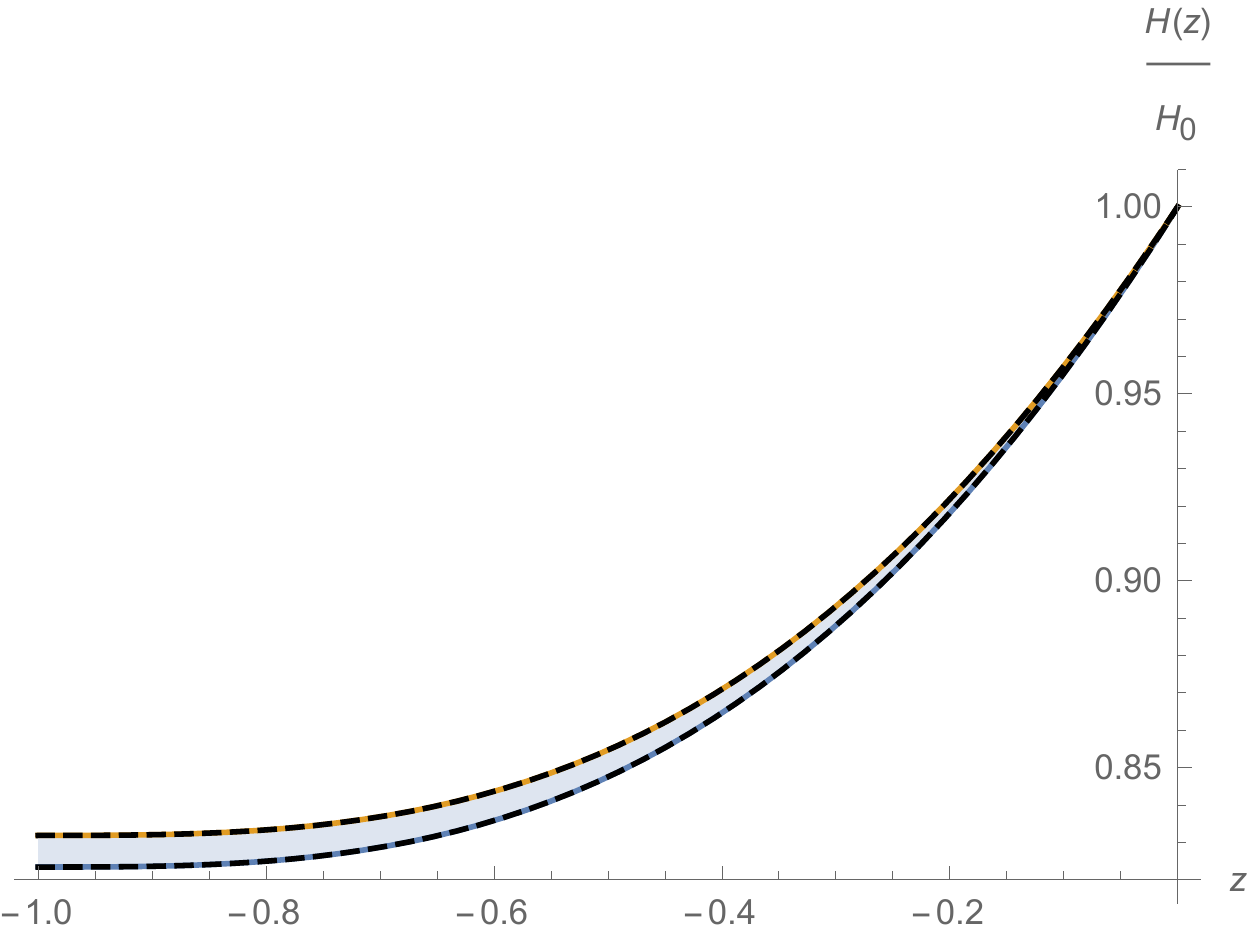}
\includegraphics[width=7.8cm,height=6cm]{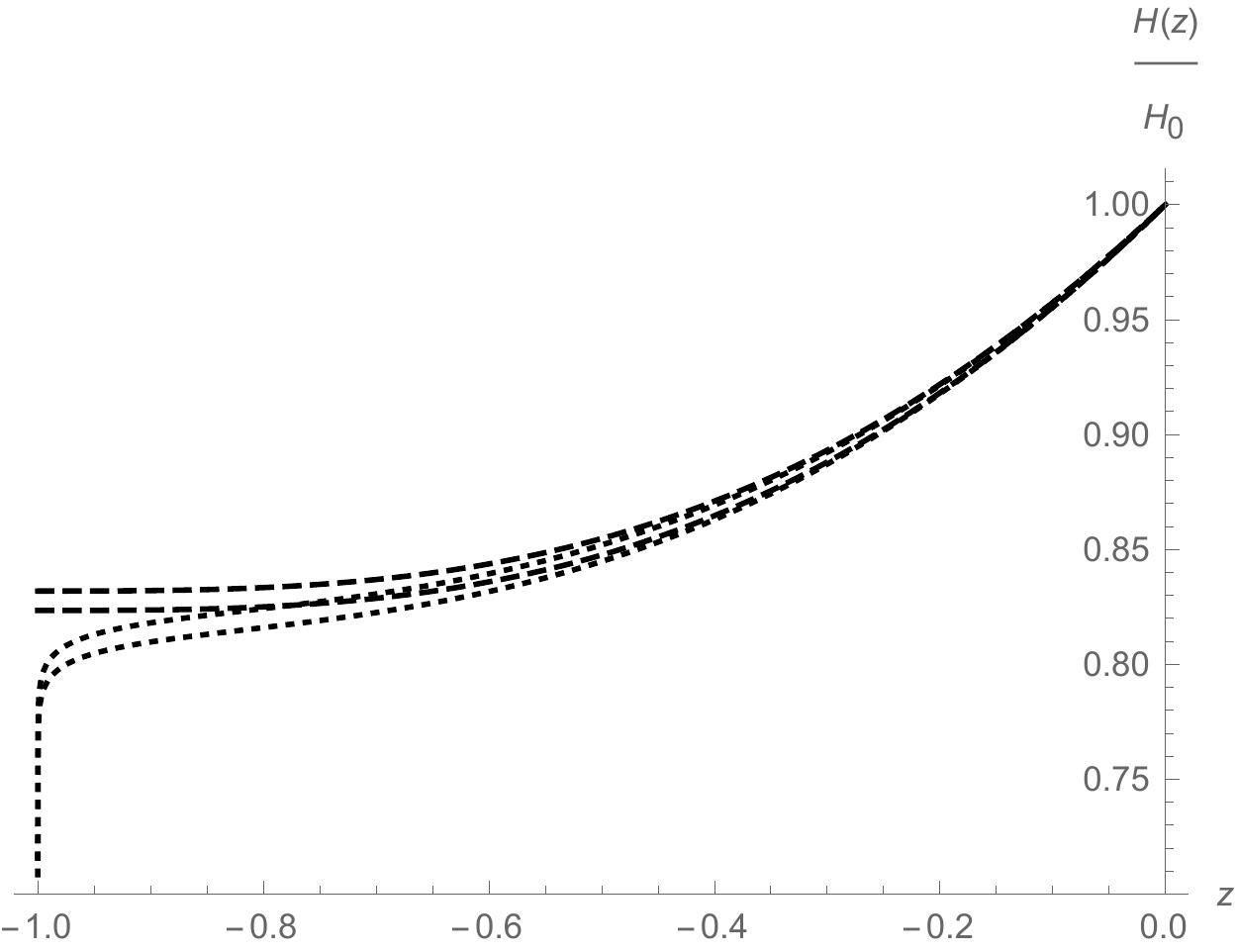}
\caption{Hubble parameter in the truncated Israel-Stewart model.}
\label{fig:hubble}
\end{figure}

In Fig. (\ref{fig:hubble}) we depict the normalized Hubble parameter, $H(z)/H_{0}$. In Ref. \cite{perturbations} the velocity of bulk viscous perturbations was constrained to the interval, $10^{-11} \ll c^{2}_{b}(\epsilon) \lesssim 10^{-8}$, in order to obtain a similar behavior as the $\Lambda$CDM model at perturbative level, therefore we will consider these values for the parameter $\epsilon$ and the interval $[10^{-6},10^{-2}]$ for the constant $\xi_{0}$. As commented before, the solution (\ref{eq:his}) depends on the value of the deceleration parameter at present time, we will consider its definition coming from the $\Lambda$CDM model $q(z) = -1 + 3(2[1+(\Omega_{\Lambda,0}/\Omega_{m,0})(1+z)^{-3}])^{-1}$ together with the normalization condition $\Omega_{\Lambda,0} + \Omega_{m,0} =1$ and the recent value reported by the Planck collaboration for $\Omega_{m,0}$ \cite{planck}. With these considerations we obtain $-0.538 \leq q_{0} \leq -0.517$. In the upper panel of Figure (\ref{fig:hubble}) we perform a comparison with the $\Lambda$CDM model (shaded region in blue) and the dashed lines corresponds to our viscous model with $\xi_{0} = 10^{-2}$, $\epsilon = 10^{-11}, 10^{-8}$ and the aforementioned values for $q_{0}$, as can be seen the model coincides with the $\Lambda$CDM model from the present time ($z=0$) to the far future ($z=-1$) and also in the past ($z>0$). In the lower panel we show the behavior of the normalized Hubble parameter when we vary the value of the constant $\xi_{0}$ to $10^{-6}$, in this case we can see that the normalized Hubble parameter tends to zero (dotted lines). As the value of the parameter $\epsilon$ grows the model deviates significantly from the $\Lambda$CDM model, but, this could represent an unstable scenario for the viscous model.      

\begin{figure}[htbp!]
\centering
\includegraphics[width=7.8cm,height=6cm]{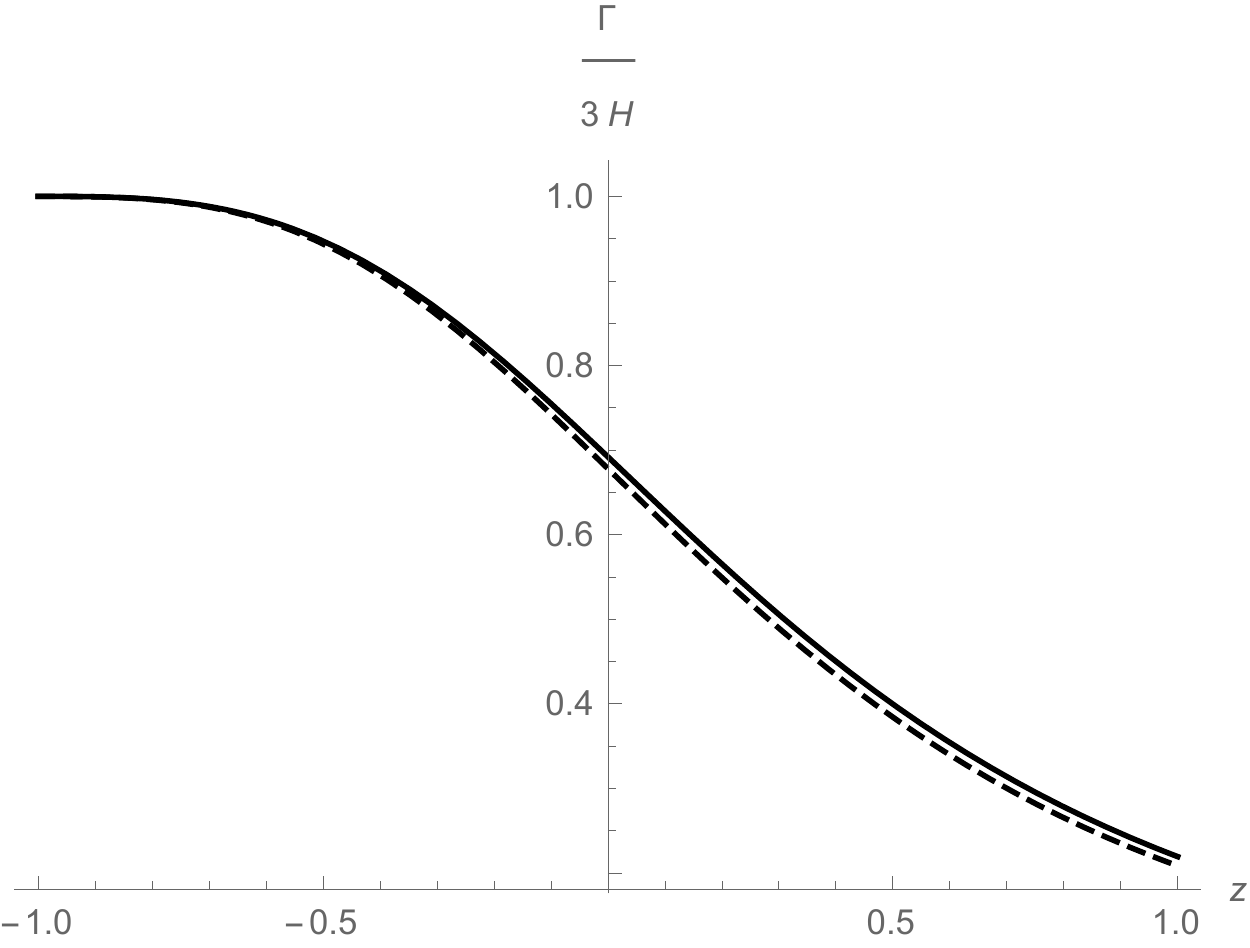}
\caption{Quotient $\Gamma/3H$.}
\label{fig:quotient}
\end{figure}

Then, if we consider the Eq. (\ref{eq:rate2}) with $\omega = 0$, we can obtain for the particle production rate
\begin{eqnarray}
    \Gamma(z) &=& 3H(z)-2(1+z)\frac{dH(z)}{dz}, \nonumber \\
    &=& H_{0}\left(1+z\right)^{\alpha}\sqrt{\frac{\cos \left(\beta \left[c_{2} + \ln \left(1+z\right) \right] \right)}{\cos \left(\beta c_{2} \right)}}\left[3-2\alpha  \right. \nonumber \\
    &+& \left. \beta \tan \left(\beta \left[c_{2} + \ln \left(1+z\right) \right] \right)\right].
\end{eqnarray}
In this case we have that if at some stage of the cosmic evolution the argument $\beta(c_{2} + \ln(1+z))$ of the tangent function takes the value $\pi(1+2n)/2$ with $n=0,\pm 1, \pm 2,...$, the particle production rate diverges. From the previous equation we can write the quotient
\begin{equation}
    \frac{\Gamma}{3H} = 1-\frac{2}{3}\alpha + \frac{\beta}{3} \tan \left(\beta \left[c_{2} + \ln \left(1+z\right) \right]\right).
\end{equation}
We show the behavior of the above expression in Fig. (\ref{fig:quotient}) with $\xi_{0} = 10^{-2}$, $\epsilon = 10^{-11}, 10^{-8}$ and $q_{0} = -0.538, -0.517$, each case is represented by the dashed and solid lines. The behavior of this quotient is related to the effective parameter as given in Eq. (\ref{eq:effective}). For $\omega = 0$ the quotient remains positive and tends to 1 at the far future ($z=-1$). This means that along the cosmic evolution we will always have dark matter production ($\Gamma > 0$) and as shown in the plot, the matter production does not become catastrophic. Relating the obtained behavior for the quotient with $\omega_{\mathrm{eff}}$ given in (\ref{eq:effective}), we can see that the cosmological viscous fluid with matter production effects from the present time ($z=0$) to the future will behave as quintessence dark energy and at the far future will emulate a cosmological constant behavior, at this stage the matter production stops. If we consider the value $\xi_{0} = 10^{-6}$ the resulting behavior for $\Gamma/3H$ is similar as the one shown in Fig. (\ref{fig:quotient}). Therefore, the truncated Israel-Stewart model with matter production does not allow a phantom scenario.  

To end this section, we discuss the following Ansatz for the Hubble parameter
\begin{equation}
    H(t > t_{s}) = \frac{\left|A\right|}{t_{s}}\left(\frac{t}{t_{s}}-1 \right)^{-1},
    \label{eq:bang}
\end{equation}
this Ansatz was proposed in Ref. \cite{bang} for the full Israel-Stewart model in order to study some of its thermodynamics properties. One interesting feature of this Ansatz is that the cosmic evolution starts from an initial singularity given at, $t = t_{s}$, that posses the characteristics of a big bang and represents an expanding universe since $\left|A\right| > 0$. Inserting the Ansatz (\ref{eq:bang}) in the truncated differential equation of the Israel-Stewart model (\ref{eq:is}), we obtain a quadratic equation for $\left|A\right|$, the solutions will be given as
\begin{equation}
    \left|A\right|_{\pm} = \frac{\left(3+3\sqrt{3}\epsilon/\xi_{0} \right) \pm \sqrt{9 + \frac{3\epsilon^{2}}{\xi^{2}_{0}}-\frac{6\sqrt{3}\epsilon}{\xi_{0}}+36\epsilon}}{3\sqrt{3}\epsilon/\xi_{0}-9\epsilon},
\end{equation}
then $\left|A\right|$ is a constant given in terms of the parameters $\epsilon$ and $\xi_{0}$. Thus, using the equation (\ref{eq:rate2}) with the Ansatz (\ref{eq:bang}), we obtain the following expression for the particle production rate
\begin{equation}
    \Gamma(t) = \frac{1}{t_{s}}\left(\frac{t}{t_{s}}-1 \right)^{-1}\left\lbrace 3\left|A\right|_{\pm}-2\right\rbrace,
\end{equation}
always that the condition $3\left|A\right|_{\pm} > 2$ it is fulfilled, we will have that the cosmic evolution starts from an initial singularity and will be driven by a dissipative fluid with infinite matter production at the beginning ($\Gamma > 0$) and such production decays as the Universe expands. In this case we have the following simple expression for the quotient between the particle production rate and the Hubble parameter
\begin{equation}
    \frac{\Gamma}{3H} = 1-\frac{2}{3\left|A\right|_{\pm}}.
\label{eq:quotient2}
\end{equation}
Adopting the values for the parameters $\xi_{0}$ and $\epsilon$ as in the previous case, we obtain some different cases for (\ref{eq:quotient2}). With $\xi_{0} = [10^{-6},10^{-2}]$ and $\epsilon = [10^{-11}, 10^{-8}]$ we have that $\Gamma/3H \approx 1$ if we consider the solution $\left|A\right|_{+}$, then by means of Eq. (\ref{eq:effective}) with $\omega = 0$, we find that the dissipative fluid with matter production behaves as a cosmological constant along the cosmic evolution. On the other hand, for $\left|A\right|_{-}$ with the same set of values for $\xi_{0}$ and $\epsilon$ we have that depending on the election of these values, the quotient can be positive but $\ll 1$, negative or zero. Using the Eq. (\ref{eq:effective}) with $\omega = 0$ we can see that for a positive quotient the effective parameter is excluded from the quintessence region and for negative quotient $\omega_{\mathrm{eff}}$ turns positive leading to a decelerated expansion. For the case of null particle production we have $q=1/2$ by means of Eq. (\ref{eq:quotient}) and this result also represents a decelerated cosmic expansion. Therefore, the interesting case is given by the solution $\left|A\right|_{+}$, but as in the case discussed previously, no phantom regime is allowed. 

\section{Non adiabatic expansion}
\label{sec:nonadia}

The results obtained in the previous sections emerge from the adiabatic condition for the entropy, i.e. $\dot{S} = 0$, leading to a constant entropy in time. However, if we consider non adiabaticity the thermodynamics description of the cosmological model becomes more consistent \cite{grandon}. From Eqs. (\ref{eq:number}), (\ref{eq:inhomo2}) and (\ref{eq:gamma}) we can compute the particle number density, yielding
\begin{equation}
    n(H) = \frac{n_{0}}{V}\left(\frac{H}{H_{0}}\right)^{2\left(1-\frac{1}{\eta}\right)},
\end{equation}
where $V$ is the Hubble volume given as $(a/a_{0})^{3}$. Note that for $H=H_{0}$, we simply have the constant density number, $n_{0}/V$, and for an expanding Universe this density is always positive. With the inclusion of dissipative effects we must consider the following evolution equation for the temperature \cite{maartens}
\begin{equation}
    \frac{\dot{T}}{T} = -3H\left(\frac{\partial p}{\partial \rho}\right) + n\dot{S}\left(\frac{\partial T}{\partial \rho}\right),
    \label{eq:temperature}
\end{equation}
the integrability condition
\begin{equation}
    \frac{\partial^{2}S}{\partial T \partial n} = \frac{\partial^{2}S}{\partial n \partial T},
\end{equation}
still holds, therefore we can write
\begin{equation}
    n\frac{\partial T}{\partial n}+(\rho +p)\frac{\partial T}{\partial \rho} = T\frac{\partial p}{\partial \rho}.
\end{equation}
If we consider these results together with Eqs. (\ref{eq:number}), (\ref{eq:entropy}) and a barotropic equation of state in the expression (\ref{eq:temperature}), we obtain
\begin{equation}
    \frac{\dot{T}}{T} = -3H\omega \underbrace{\frac{1 + \frac{3\Pi}{\rho(1+\omega)}+\frac{\Gamma}{3H}}{1 + \frac{\Pi}{\rho(1+\omega)\left( 1-\frac{\Gamma}{3H}\right)}+\frac{\Gamma}{3H\left( 1-\frac{\Gamma}{3H}\right)}},}_{\Xi(t)}
\end{equation}
by considering the relationship between the redshift and the scale factor the equation for the temperature can be written as
\begin{equation}
    \frac{1}{T}\frac{dT}{dz} = 3\omega(1+z)^{-1}\Xi(z),
\end{equation}
if we integrate
\begin{equation}
    T(z) = T_{0}\exp \left[3\omega\int^{z}_{0}\frac{\Xi(z')}{1+z'}dz'\right]. 
\end{equation}
It is worthy to mention that the obtained expression for the temperature is definite positive, in the case where we have null contribution from dissipative and matter creation effects, $\Pi = \Gamma = 0$, the temperature takes the standard definition obtained in a single fluid description, $T(z) = T_{0}(1+z)^{3\omega}$ and for the standard dark matter case the temperature takes a constant value, as in the $\Lambda$CDM model.\\ 

On the other hand, if we consider the expression (\ref{eq:entropy}), we can write
\begin{equation}
    \dot{S} = \frac{9H^{3}}{nT}\left[-\frac{2}{3}(1+q)+(1+\omega)\left(1-\frac{\Gamma}{3H}\right)\right],
\end{equation}
and the expression (\ref{eq:viscpress}) was considered for the bulk viscous pressure. Now, inserting Eq. (\ref{eq:quotientnonadia}) in the previous result one gets
\begin{equation}
    \dot{S} = \frac{9H^{3}}{nT}\left\lbrace \frac{2}{3}(1+q)\left[(1+\omega)\left(1-\frac{1}{\eta} \right) - 1\right] + (1+\omega)\right\rbrace.
\end{equation}
As can be seen the entropy production has contributions from matter creation and dissipative effects. For the standard dark matter case we have
\begin{eqnarray}
\dot{S} =\left\{ 
\begin{array}{c}
\frac{9H^{3}}{nT_{0}}\left\lbrace 1-\frac{2}{3}\frac{(1+q)}{\eta}\right\rbrace, \ \mbox{quintessence},
\\ 
\frac{9H^{3}}{nT_{0}}\left\lbrace 1+\frac{2}{3}\frac{\left|1+q\right|}{\eta}\right\rbrace, \ \ \ \ \  \mbox{phantom},
\label{eq:dotS}
\end{array}
\right. 
\end{eqnarray}
where we have considered that at effective level both cases can appear. In order to be in agreement with second law of thermodynamics, $\dot{S} > 0$, the conditions $T_{0} > 0$ and $n > 0$, must be satisfied together with the following cases: (i) in the quintessence scenario the conditions $1 > 2(1+q)/3\eta$ and $\eta > 0$ must be satisfied, the positivity of the entropy production is guaranteed for $\eta < 0$, (ii) for the phantom regime the fulfillment of the second law is achieved with $\eta > 0$ and for $\eta < 0$ we must have $1 > 2\left|1+q\right|/3\eta$. Thus, under the non adiabatic condition for the entropy the cosmic expansion is free of the negative entropy or temperature problem \cite{chemical}.   

\section{Final remarks}
\label{sec:final}

In this work we studied the cosmic evolution that emerges from the consideration of matter creation effects in a viscous fluid under the adiabatic condition for the entropy, i.e., constant entropy. As first approach we adopted an inhomogeneous Ansatz for the particle production rate, $\Gamma$. In this first scheme we obtained that the model at effective level could describe phantom or quintessence regimes depending on the values of the parameters $\eta$ and $\omega$, being the case of interest the dark matter type behavior for created matter denoted by $\omega = 0$. For this special case phantom and quintessence behaviors are still present. On the other hand, we computed the dimensionless age of this kind of Universe for two cases: created matter represents the whole content of Universe and secondly created matter represents the actual matter of the Universe; for both cases we got that the age for these kind of universes deviates from 1 but remains close to this value, this is characteristic of universes with presence of dark energy.  However, when the deceleration parameter is computed in this first description we obtain a constant parameter as result, therefore a model of this kind it is not consistent with the cosmological observations or the $\Lambda$CDM model. In order to fix this issue we considered the inclusion of baryonic matter and performed the statistical analysis for the model in order to constraint the parameters $\Omega_{B}$ and $\eta$ with $\omega = 0$. Therefore, with the constrained parameters we proved that this model behaves as quintessence dark energy along the cosmic evolution.\\

As second approach we dropped the Ansatz philosophy for the particle production rate and we constructed such term with the consideration of two descriptions for the bulk viscous pressure: the Eckart model and the truncated version of the Israel-Stewart formalism. The particle production rate has a relevant role in the cosmic evolution since
\begin{equation*}
    \omega_{\mathrm{eff}} = - \frac{\Gamma}{3H},
\end{equation*}
for $\omega = 0$, as discussed previously. If $\Gamma > 3H$, the phantom scenario is allowed and the case $\Gamma < 3H$ could represent quintessence or decelerated expansion. In the Eckart description a big rip singularity appears, then in this case the $\Gamma$ term also exhibits a singular behavior leading to a catastrophic matter production. This scenario contradicts the expected behavior of the phantom regime where matter or any structure must be diluted. We infer that this conduct is due to the fact that the Eckart formalism is non-causal, therefore it is not appropriate to describe late times in any type of Universe. On the other hand, if the truncated version of the Israel-Stewart model is considered we have several possibilities to study, however, in this work we focused only on two cases: an analytical solution for the Hubble parameter emerging from this approach and an Ansatz for the Hubble parameter that leads to a cosmic evolution with an initial singularity with the properties of a Big Bang. It is worthwhile to mention that in this scenario the catastrophic matter production is not present. For the analytical solution of the Hubble parameter we obtained that the corresponding $\Gamma$ term leads to a quintessence dark energy evolution and at the far future the model could mimic the $\Lambda$CDM model. These scenarios are possible if the parameter $\epsilon$ (a constant parameter responsible of characterizing the velocity of bulk viscous perturbations) lies in a specific interval such that at perturbative level the viscous model is close to $\Lambda$CDM, if we consider other values for $\epsilon$ we could have other possibilities for the cosmic evolution of the model but our description could reveal an unstable behavior. When the Ansatz for the Hubble parameter is considered, we have two branches characterized by $\left| A\right|_{\pm}$; for the branch $\left| A\right|_{+}$ the cosmological constant expansion is emulated by our model and if we consider the branch $\left| A\right|_{-}$ decelerated expansion is obtained. Then, this latter case is not favored and must be discarded.\\

We must emphasize that it seems that the scenario of dark matter type production plus other cosmological effects does not allow the crossing to the phantom regime, i.e, we only have access to a quintessence dark energy scheme. The case of dark matter type production was discussed by the authors in Ref. \cite{us1}. We would like to comment that according to some recent tentative results the quintessence scenario seems to be the elected candidate by the observable Universe to lead its late times behavior, in Ref. \cite{prllast} it is reported that the cosmic expansion could be slower than we think since the catalyst of such expansion is a quintessence component. Thus, the kind of model as studied in this work could provide a viable theoretical framework to describe our Universe.\\

Finally, we discuss the cosmic expansion of our model if we implement the non adiabaticity condition. In this description we obtain that the temperature of the viscous fluid is positive definite and in the case of dark matter type behavior, $\omega=0$, we got that such temperature takes a constant value as in the $\Lambda$CDM model even if $\Pi$ and $\Gamma$ are different from zero. For $\Pi = \Gamma = 0$, we recover the single fluid description at thermodynamics level. Then, the model obeys the second law if we take into account some conditions for the parameters of the model. These phantom or quintessence scenarios have positive entropy and temperature.

\section*{Acknowledgments}
M.C. work was supported by S.N.I. (CONACyT-M\'exico).

\end{document}